# Enhancement and Tunability of Near-Field Radiative Heat Transfer Mediated by Surface Plasmon Polaritons in Thin Plasmonic Films

Svetlana V. Boriskina *, Jonathan K. Tong, Yi Huang, Jiawei Zhou, Vazrik Chiloyan, Gang Chen *

Department of Mechanical Engineering, Massachusetts Institute of Technology, Cambridge, MA 02139, USA

* Authors to whom correspondence should be addressed; E-Mails: sborisk@mit.edu (S.V.B.); gchen2@mit.edu (G.C.)

**Abstract:** The properties of thermal radiation exchange between hot and cold objects can be strongly modified if they interact in the near field where electromagnetic coupling occurs across gaps narrower than the dominant wavelength of thermal radiation. Using a rigorous fluctuational electrodynamics approach, we predict that ultra-thin films of plasmonic materials can be used to dramatically enhance near-field heat transfer. The total spectrally integrated film-to-film heat transfer is over an order of magnitude larger than between the same materials in bulk form and also exceeds the levels achievable with polar dielectrics such as SiC. We attribute this enhancement to the significant spectral broadening of radiative heat transfer due to coupling between surface plasmon polaritons (SPPs) on both sides of each thin film. We show that the radiative heat flux spectrum can be further shaped by the choice of the substrate onto which the thin film is deposited. In particular, substrates supporting surface phonon polaritons (SPhP) strongly modify the heat flux spectrum owing to the interactions between SPPs on thin films and SPhPs of the substrate. The use of thin film phase change materials on polar dielectric substrates allows for dynamic switching of the heat flux spectrum between SPP-mediated and SPhP-mediated peaks.

## 1. Introduction

Radiative heat transfer between surfaces separated by distances comparable to or shorter than the wavelength of thermal radiation is well-known to deviate from the classical Planck's law of the blackbody (BB) radiation [1–7]. Overall, the total heat flux between two surfaces depends on the number of channels available for the energy transfer in both photon energy and momentum space, while the capacity of each channel is governed by the temperatures of the hot and cold sides [7]. Surface waves propagating along the material–vacuum interfaces [4–6,8–10] can dramatically increase the number of available channels within a certain wavelength range, thus playing a major role in enhancing and spectrally-tailoring near-field radiative heat transfer across sub-micron gaps. In particular, near-field heat transfer mediated by surface phonon polariton (SPhP) waves supported in silicon carbide (SiC) and silica (SiO$_2$) have been extensively studied and shown to increase radiative heat transfer by orders of magnitude beyond the predictions from Planck's law [3,4,9,11–14]. Other materials supporting SPhP modes have also been explored, including vanadium dioxide (VO$_2$) in its insulator phase [15] and ferroelectrics such as perovskites [16]. The enhancement and spectral-tunability of heat exchange between hot and cold surfaces may benefit applications including on-chip heat management [17], heat-assisted magnetic recording [18], nanofabrication [19], thermal imaging [20], and thermophotovoltaic (TPV) energy conversion [21–27].

Another type of surface waves supported by material–vacuum interfaces are surface plasmon polaritons (SPP), which result from the coupling between collective oscillations of free charge carriers in materials (*i.e.*, surface plasmons) and electromagnetic waves in vacuum (or dielectrics) [28,29]. SPP waves can propagate along the surfaces of materials with a high concentration of free charge carriers, such as metals and highly-doped semiconductors [30–35]. Similar to SPhP modes, SPP modes exhibit a high local density of optical states (DOS), which strongly modifies the intensity and spectral shape of the radiative heat flux across narrow gaps. Conventional metals, however, are typically considered poor candidates for near-field heat transfer [36–38]. Despite earlier theoretical predictions that high-resistivity metals are optimal in maximizing near-field transfer, at least in the limit of infinitesimal gaps [7], calculations and measurements of the heat flux across non-vanishing gaps show that dielectrics supporting high-DOS SPhP waves significantly outperform conventional metals such as gold, aluminum, and tungsten.

One reason is the poor spectral overlap of SPP modes with the Planck's distribution even at temperatures as high as 1000 K. According to Wien's displacement law, the peak wavelength for thermal emission is $\lambda_T = b \times T^{-1}$ where $b = 2898$ $\mu m \times K$. Although $\lambda_T$ shifts from 9.7 μm at 300 K to 2.9 μm at 1000 K, it is still much longer than the SPP wavelengths of most conventional metals, which typically occur in the ultra-violet to near-infrared wavelength range. As a result, for conventional metals, the major contribution to the near field energy flux does not come from the coupled SPP modes, but rather from the evanescent fields resulting from the total internal reflection of low-energy photons [36]. By comparison, the SPhP resonances of many polar materials, such as SiO$_2$, occur within the mid- and far-infrared wavelength range. However, SPhP modes are intrinsically constrained by the material's atomic bonding, while SPP modes can be tuned by their plasma frequency to span from the ultra-violet to the far infrared wavelength range based on the concentration of free carriers [39]. This wide SPP wavelength range is a major advantage of plasmonic materials over polar dielectrics supporting SPhP waves. Recently, many new plasmonic materials with plasma frequencies either in the visible or near-infrared frequency bands were explored. These materials are promising candidates for near-field heat transfer enhancement, especially at high temperatures relevant to energy-harvesting applications such as e.g., TPV. These include metal oxides and nitrides [40–43], graphene [32,34,44,45], and highly doped semiconductors such as Si [30,31,33,46,47], InSb and InAs [48]. In this study, we will focus on metal oxides as an emerging interesting group of plasmonic materials with near-infrared plasma frequencies.

The other reason why plasmonic materials do not perform as well as polar dielectrics for near-field heat transfer applications is the higher level of their dissipative losses, which dampen SPP mode resonances [49]. Therefore, despite numerical predictions that ideal doped semiconductors with optimally-chosen material parameters can exhibit a maximum SPP-mediated heat transfer that exceeds the maximum SPhP-mediated one [50], amongst the real bulk plasmonic materials, the near-field radiative heat flux typically falls below that achievable with polar dielectrics.

Near-field radiative heat transfer can be further enhanced by the change of the emitter and absorber morphology from isotropic bulk material to a multi-layer artificial metamaterial [51–55] or even a single-layer thin film [56,57]. Here, we numerically demonstrate that low-dimensional emitters and absorbers supporting SPP modes can lead to dramatic increases in near-field heat transfer, with ultra-thin metal films *outperforming their bulk counterparts and SiC emitters/absorbers by over an order of magnitude* in terms of the total spectrally integrated heat flux. This increase is significantly more pronounced than the one observed in thin-film polar dielectrics [57,58]. We use a rigorous electrodynamics approach based on the fluctuation–dissipation theorem [59] to perform a comprehensive investigation of the role of the film thickness, metal dissipative losses, and substrate effects on the efficiency and spectral characteristics of the near-field heat transfer. We identify the best existing and prospective material candidates to maximize high-temperature and high-frequency heat transfer, which could prove useful in emerging applications such as near-field thermophotovoltaics [21–23,25,35,43]. We also demonstrate *spectrally-tailored and dynamically-tunable heat transfer* between compound multi-layered emitters and absorbers made of phase-change materials that support both SPP and SPhP modes within partially overlapping frequency ranges.

## 2. Near-Field Radiative Heat Transfer Mediated by the Coupling of Surface Polariton Waves

Thermal radiation fundamentally originates from thermally-induced dipole fluctuations caused by the chaotic thermal motion of electrons and atoms in the emitter material, which generate electromagnetic fields. These fluctuations lead to the excitation of both propagating and evanescent waves with both types of waves being able to carry energy between the hot and the cold surfaces. A schematic illustration detailing the mechanism of heat exchange between two surfaces via evanescently-coupled SPP waves is shown in Figure 1a and is contrasted with alternative mechanisms of radiative heat transfer via propagating waves (Figure 1b) and via evanescent fields created by the mechanism of the total internal reflection (Figure 1c).

Within the framework of fluctuational electrodynamics [59], the heat flux between two parallel plates in the configuration shown in Figure 1 can be formulated as follows:

$$Q_{1-2} = \frac{1}{4\pi^2} \int_0^\infty (f(\omega, T_1) - f(\omega, T_2)) d\omega \cdot \sum_{j=s,p} \int_0^\infty \mathcal{T}_j(\omega, k) k \, dk \quad (1)$$

where $f(\omega,T) = \hbar\omega \cdot (\exp(\hbar\omega/\kappa_B T) - 1)^{-1}$ is the mean energy of the Planck oscillator at frequency $\omega$ in thermal equilibrium at temperature $T$ without the zero point energy term $\hbar\omega/2$, which cancels in Equation (1), $\hbar$ is Planck's constant, $\kappa_B$ is Boltzmann constant, and $k = k_\parallel$ is the wavevector component parallel to the interface. Function $f$ suppresses the contribution to the heat flux from the channels with frequencies much larger than $\kappa_B T/\hbar$. The transmission coefficients for s- and p-polarized radiative ($k \leq k_0$) and evanescent ($k > k_0$) waves contributing to the total heat transfer have the following form [3,9,24]:

$$\mathcal{T}_j = \begin{cases} \left(1-\left|r_{01}^j\right|^2\right)\cdot\left(1-\left|r_{02}^j\right|^2\right)\Big/\left|1-r_{01}^j r_{02}^j e^{2ik_{\perp 0}d}\right|^2, & k \leq k_0 \\ 4e^{-2\,\text{Im}(k_{\perp 0})d}\cdot \text{Im}(r_{01}^j)\cdot \text{Im}(r_{02}^j)\Big/\left|1-r_{01}^j r_{02}^j e^{-2\,\text{Im}(k_{\perp 0})d}\right|^2, & k > k_0 \end{cases} \quad (2)$$

where $k_0 = \omega/c$ is the free space wavevector, $k_{\perp m} = \left(\varepsilon_m k_0^2 - k^2\right)^{1/2}$ are the wavevector components in different media ($m$=0,1,2) perpendicular to the surfaces, and $r_{0m}^s$, $r_{0m}^p$ are the polarization-dependent Fresnel reflection coefficients at the interface separating medium 0 (vacuum gap) and medium $m$ ($m$=1,2). In the simplest case of the bulk emitter and absorber, $r_{0m}^s = (k_{\perp 0} - k_{\perp m})/(k_{\perp 0} + k_{\perp m})$, $r_{0m}^p = (\varepsilon_m k_{\perp 0} - \varepsilon_0 k_{\perp m})/(\varepsilon_m k_{\perp 0} + \varepsilon_0 k_{\perp m})$. Reflection coefficients for more general cases of multi-layered emitters and absorbers are shown in the Supplementary materials. As already noted, two types of surface waves can contribute to the total heat flux: totally internally reflected waves with $k_0 < k < k_0|\sqrt{\varepsilon_m}|$ and surface polariton waves with high parallel momentum values $k > k_0|\sqrt{\varepsilon_m}|$. In particular, if both surfaces can support polariton waves, the waves excited on each surface resonantly couple forming bonding and anti-bonding supermodes [28,60,61], which strongly modifies resonant photon coupling across the vacuum gap.

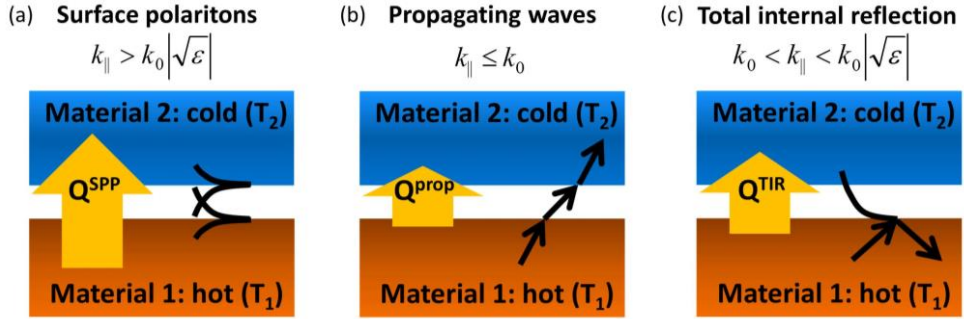

**Figure 1.** A schematic of the near-field radiative heat exchange between a parallel-plate emitter (medium 1) and absorber (medium 2) across the vacuum gap (medium 0). Heat transfer between the hot and cold sides mediated by (**a**) evanescently-coupled surface polariton modes (if supported by the materials), (**b**) propagating waves in the vacuum gap, and (**c**) evanescent fields created by the total internal reflection of propagating waves within the emitter material.

To better understand the physical mechanisms that distinguish SPP modes from SPhP modes, we begin by comparing the dielectric functions of metals and polar dielectrics. The dielectric constants of plasmonic materials can be approximated via the free-electron Drude model:

$$\varepsilon_D(\omega) = \varepsilon_\infty - \frac{\omega_p^2}{\left(\omega^2 + i\gamma\omega\right)} \quad (3)$$

where $\varepsilon_\infty$ is the high-frequency permittivity limit, $\omega_p^2 = ne^2/m_e\varepsilon_0$ is the plasma frequency and the damping term $\gamma = e/\mu m_e$ is the collision frequency, which are functions of the charge carrier density $n$, the charge of the electron $e$, the vacuum permittivity $\varepsilon_0$, the effective electron mass $m_e$, and the electron mobility in the material $\mu$.

For polar dielectrics, the dielectric permittivities can be described via the Lorentz oscillator model as follows:

$$\varepsilon_L(\omega) = \varepsilon_\infty \left( 1 + \sum_{(j)} \frac{\omega_{j,LO}^2 - \omega_{j,TO}^2}{\omega_{j,TO}^2 - \omega^2 - i g_j \omega} \right) \quad (4)$$

where $\omega_{j,TO}^2$ and $\omega_{j,LO}^2$ are the frequencies of the $j^{\text{th}}$ transverse and longitudinal optical phonon modes, respectively, and $g_j$ is the damping constant.

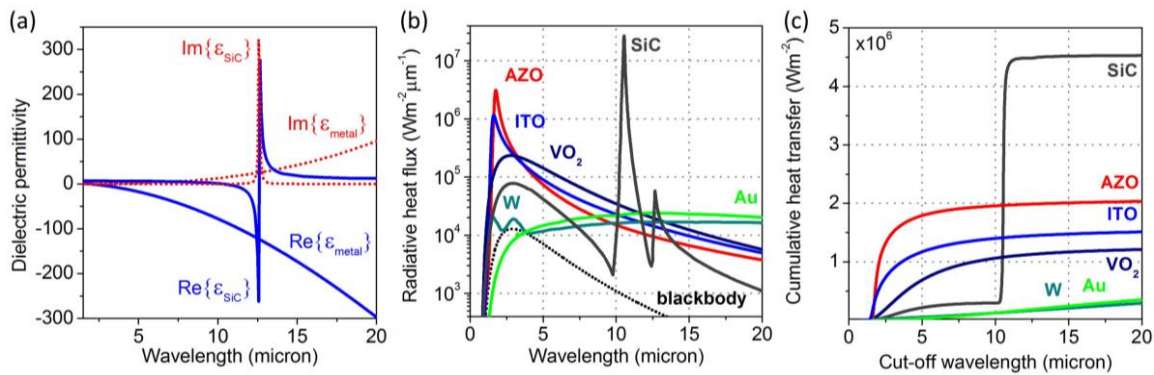

**Figure 2.** SPP- and SPhP-mediated heat flux spectra. (**a**) Complex dielectric permittivities of the polar dielectric (SiC) and a Drude-like metal with $\lambda_p = 1.1 \mu m$, $\gamma = 30 THz$, $\varepsilon_\infty = 3$ (blue solid lines: real parts, red dotted lines: imaginary parts). (**b**) Heat flux spectra as a function of wavelength between semi-infinite bulk emitters and absorbers supporting surface polaritons. The emitter and absorber are separated by a 20 nm gap. The hot and cold side temperatures are 1000 K and 360 K, respectively. The 360 K value was chosen to ensure that VO$_2$ is in its metallic form [62]. The blackbody heat flux is shown as the dotted line for comparison. Au, SiC and W permittivity values are taken from [63], the ITO and AZO ones from [64], and VO$_2$ values from [62]. (**c**) Cumulative heat flux per unit area in the spectral range below a cut-off wavelength $\lambda_c$ between the same materials as in (**b**).

Polar dielectrics with low damping parameters $g_j$ feature $j$ bands of the negative real part of the dielectric permittivity in the frequency ranges roughly between $\omega_{j,LO}$ and $\omega_{j,TO}$. On the contrary, the real part of the dielectric permittivity for a plasmonic material is negative for a wide range of frequencies lower than the material plasma frequency. The two types of dielectric functions are compared in Figure 2a, which shows the permittivities of SiC and a sample Drude-like metal with parameters typical for infrared plasmonic materials such as indium tin oxide (ITO), aluminum-doped zinc oxide (AZO), and so on [64]. Polar dielectrics have much smaller dissipative losses than metals, which result in flatter dispersion characteristics that reach larger in-plane wave vectors. Therefore, lower losses result in a higher DOS and thus a higher resonant SPhP-mediated near-field energy flux. In the limit of high in-plane photon momentum, the heat flux scales inversely with the material dissipative losses and is proportional to $\exp(-2kd)(4/\text{Im}\varepsilon)^2$ [49]. This scaling law is, however, limited to the cases when the radiative heat flux is dominated by the contribution from the very large k-vectors at the strong polariton-enhanced peak. In general, a tradeoff between the peak(s) amplitude and width dictates the optimum level of material dissipative losses. This is illustrated in Figure 2b, which shows the near-field radiative heat flux spectra mediated by SPP waves for different plasmonic materials with a comparison to the well-studied case of near-field heat transfer via coupled SPhP waves on SiC-air interfaces. For this comparison, we chose different types of plasmonic materials, including noble (Au) and transitional (W) metals as well as metal oxides (ITO, AZO, and $VO_2$). SiC was chosen as a benchmark material, as it is often used as an etalon in the estimates of the near-field heat transfer enhancement [65]. As shown, the peak value of the heat transfer function between a bulk SiC emitter and absorber exceeds the values achievable with any of the plasmonic materials even those whose plasma frequencies provide better overlap with the blackbody spectra.

The frequency of the peak heat transfer via coupled SPPs on metal–vacuum interfaces can be approximately calculated by setting $\varepsilon(\omega) = -1$ in Equation (3), resulting in the following expression:

$$\omega_{peak} = \left( \frac{\omega_p^2}{\varepsilon_\infty + 1} - \gamma^2 \right)^{1/2} \quad (5)$$

SPP modes in conventional noble and transitional metals such as gold (Au), tungsten (W) overlap poorly with the thermal emission spectrum owing to their Drude frequencies being in the ultraviolet or visible spectral regions. Accordingly, the meta-metal heat flux (shown in Figure 2b) is not significantly enhanced over the blackbody limit, and the flux spectra do not feature well-defined peaks corresponding to the coupled SPP waves. In contrast, the heat-transfer spectra between some metal oxides whose SPP modes overlap well with the thermal spectrum (AZO and ITO) feature pronounced peaks at the wavelengths defined by Equation (5). However, due to higher level of dissipative losses, these peaks have a lower peak intensity compared to SiC, but are much broader. Vanadium dioxide ($VO_2$) also has a broad SPP peak, which is blue-shifted from the thermal emission spectrum.

Figure 2c further supports the above observations by visualizing the accumulated heat flux transferred by photons with energies above sliding cut-off energy. The plots in Figure 2c clearly show how the SPhC-mediated heat transfer between SiC interfaces is achieved via photons within a narrow spectral range, while the SPP-mediated heat exchange between plasmonic materials is carried by photons across a wide range of energies. This leaves room for optimization of the total spectrally-integrated heat flux. In fact, some lossy plasmonic materials can outperform less lossy materials by compensating the reduction in the peak flux amplitude with an increase in the SPP peak frequency bandwidth [39,43]. Plasmonic materials with dielectric permittivities deviating from the Drude model predictions also demonstrate additional peaks in their near-field heat transfer spectra, as evident for tungsten in Figure 2b. Nevertheless, the high dissipative losses in plasmonic materials are overall a limiting factor to their performance. As a result, the total integrated heat fluxes for all the plasmonic materials considered here fall below the SiC benchmark (Figure 2c).

## 3. Near-Field Heat Transfer Increase by SPP Mode Splitting in Ultra-thin Plasmonic Films

Despite the detrimental effect of the dissipative losses on the near-field heat flux between SPP-supporting surfaces, plasmonic materials offer two significant advantages over polar dielectrics.

First, the plasma frequency in materials with free charge carriers is tunable by varying the carrier density [39]. This enables synthesis of materials with heat transfer peaks tailored to certain frequency ranges for various operational temperatures. Second, the wide frequency range of negative permittivity values offers a large design space where SPP modes can exist. The dispersion characteristics of these modes can be modified by changing the morphology of the emitter and/or absorber rather than their material composition. In particular, we propose using ultrathin films of plasmonic materials to strongly enhance the SPP-mediated near-field energy transfer across vacuum gaps.

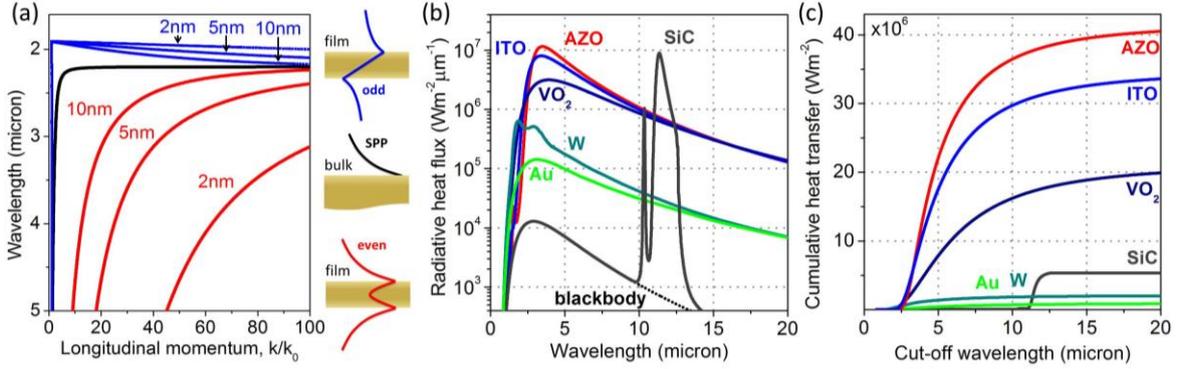

**Figure 3.** SPP-mediated heat transfer between ultrathin films. (**a**) The band structure of SPP modes for a single metal-air interface (black) and thin films with progressively smaller thicknesses. The red and blue branches correspond to the even and odd coupled SPP modes, respectively. Schematics of the modal field profiles are shown in the insets. The Drude model parameters are $\lambda_p = 1.1 \mu m$, $\gamma = 0$, $\varepsilon_\infty = 3$. (**b**) The heat flux spectra as a function of wavelength between 2 nm-thin emitters and absorbers supporting surface polaritons. The gap width is 20 nm, and the hot and cold side temperatures are 1000 K and 360 K, respectively. The blackbody heat flux is shown as the dotted line for comparison. (**c**) Cumulative heat flux per unit area in the spectral range below a cut-off wavelength $\lambda_c$ between the same materials as in (**b**).

The mechanism to enhance heat transfer is based on the splitting of modal branches in the SPP dispersion owing to the electromagnetic coupling of the SPP modes on the opposite sides of the thin films as illustrated in Figure 3a. The two branches correspond to odd and even coupled SPP modes, whose frequency splitting increases as the metal film thickness decreases. This simple tuning of the dispersion via morphology dramatically increases the frequency range where high-momentum SPP states exist and contribute to the near-field radiative heat transfer (see Figure 3b,c and Figure 4). It should be noted that the SPhP modes in polar dielectrics experience the same changes in their band structure when the emitter thickness is reduced [12,57,66–70]. However, as the SPhP modes only exist within a narrow frequency band between the longitudinal and transverse optical phonon frequencies, the modal splitting and coupling does not significantly affect the heat flux across the gap.

Figures 3b and 3c show the heat flux spectra (b) and accumulated flux spectra (c) between thin films of the same materials as in Figure 2b,c. A comparison between the two figures reveals that plasmonic films provide over an order of magnitude enhancement over their bulk counterparts, while SiC films provide only a slight increase to the total spectrally integrated heat flux. As a result, the

SPP-enhanced heat transfer between metal films with plasma frequencies in the near-IR range becomes significantly larger than that between SiC films. It should also be noted that a large portion of the energy exchanged between plasmonic films is being carried by higher-energy near-IR photons, which may be useful for some applications such as TPV.

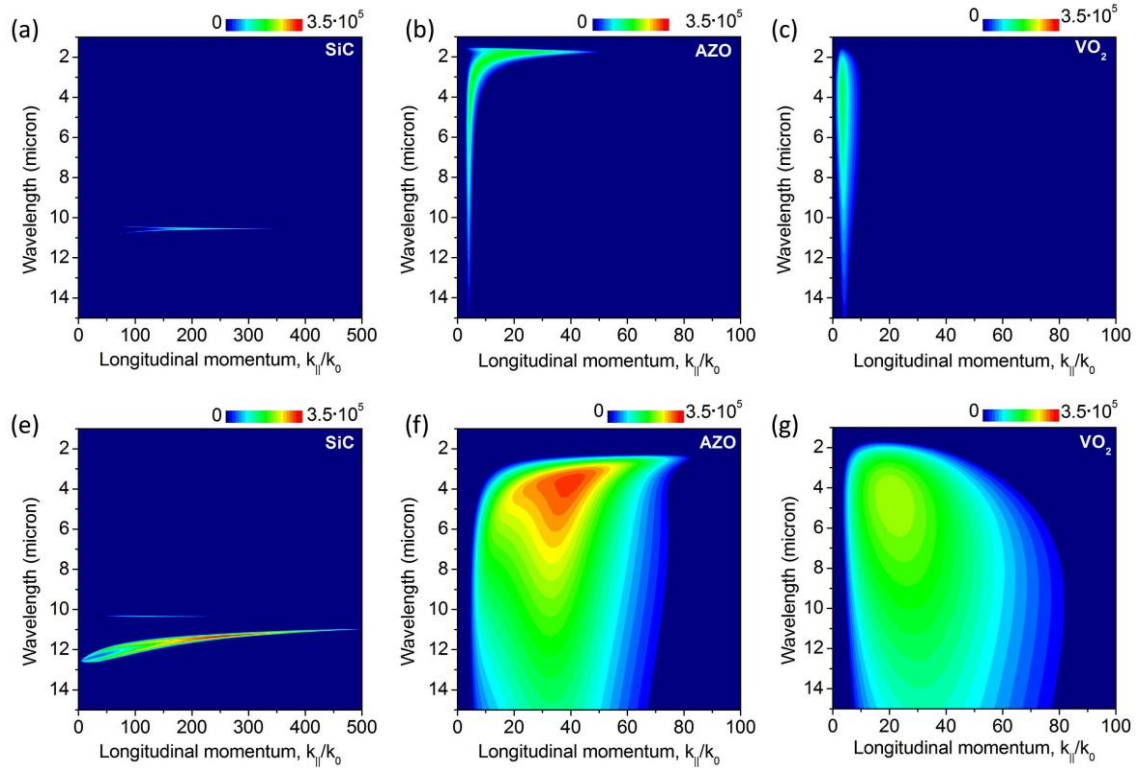

**Figure 4.** Wavevector-resolved transmission coefficients from Equation (2) for bulk (**a**–**c**) *versus* thin 2 nm-thin film (**e**–**g**) energy transfer for (**a,e**) SiC, (**b,f**) AZO, and (**c,g**) Vanadium dioxide (VO$_2$). The emitter and absorber are separated by a 20 nm gap, and the hot and cold side temperatures are 1000 K and 360 K, respectively.

The differences in near-field heat transfer between SiC films and plasmonic films can be revealed by inspecting the wavevector-resolved transmission coefficients defined by Equation (2), as shown in Figure 4. Here, we chose AZO and VO$_2$ to illustrate two characteristic cases of the SPP-mediated heat transfer: (i) with the SPP peak overlapping the thermal emission spectrum (AZO), and (ii) with the SPP peak blue-shifted from the thermal emission spectrum (VO$_2$). It can be seen in Figure 4a that the peaks in the heat flux spectrum between SiC emitters correspond to the high-momentum states associated with the excitation of the SPhP waves within a narrow frequency band. This band widens somewhat in the case of the thin film emitter, which translates into a higher heat flux, as seen in Figures 2b and 3b. However, in the case of plasmonic emitters, photon states within a broad range of frequencies contribute to the heat flux. In the momentum space, the largest-k contributing channels for heat transfer are determined by the exponential decay term in Equation (2), $\exp(-2\,\mathrm{Im}(k_{z0})d)$, which suppresses channels with $k >> (2d)^{-1}$. Owing to the lower dissipative losses, this characteristic momentum is much higher for SiC emitters than for the plasmonic ones when comparing Figure 4a,e with Figure 4b,c,f,g. However, the width of the spectral peaks for SPP-mediated heat flux in the frequency domain compensates for the suppression of the high-momentum states.

## 4. Optimization of the Plasmonic Films Materials and Morphology

For radiative heat transfer between two films, the number of channels available for energy exchange is determined by both the film material and its morphology (*i.e.*, thickness), as both factors modify the number, intensity and bandwidth of the SPP resonances. Accordingly, it is important to estimate the optimum film parameters to achieve optimal coupling between emitting surfaces that would maximize the radiative heat flux. Previous work estimated that for either SPhP or SPP-mediated film-to-film heat transfer, there is an optimum film thickness that provides the highest rate of near-field transfer in the near-field regime [57]. Asymptotic analytical approximations predict that the layers should not be too thin, with a thickness that is inversely proportional to the absolute value of the permittivity at the surface polariton resonance [57]. However, our rigorous computations, based on fluctuational electrodynamics, reveal that there is a fundamental difference in the film thickness optimization between polar dielectrics and plasmonic materials.

This difference is illustrated in Figure 5a, which shows how the total spectrally integrated heat flux scales with the film thickness. In the case of SiC films, a local maximum can be observed at film thickness of about 20 nm, while the heat flux between plasmonic films progressively increases with the reduced film thickness. The observed difference stems from the different frequency ranges where the surface polariton modes can exist. The limited bandwidth available for SPhP modes of SiC results in the optimum mode splitting that maximizes the heat flux. In contrast, for plasmonic materials, the negative-permittivity window is not bound on the low energy side. Thus, a larger splitting of SPP branches always increases the total heat flux. Accordingly, within the framework of the classical fluctuational electrodynamics with local dielectric permittivities, the optimal film thickness for a plasmonic material is the thinnest achievable for a given material and fabrication procedure. Additional factors may arise that affect the maximum achievable radiative heat flux between ultrathin films (see detailed discussion in Section 6). However, according to the data shown in Figure 5a, the enhancement over the bulk case already becomes pronounced at film thicknesses at and below 30 nm. Furthermore, the SiC limit can be surpassed with films of thicknesses about 20 nm and below, which are well within the limits of classical approximations used in this study.

A comparison of the heat flux spectra for SiC and plasmonic films in Figure 3b shows that the SPP peaks for the plasmonic films that outperform SiC are better aligned with the peak in the Planck's distribution at 1000 K, which is centered at 2.9 μm. To estimate how these materials will perform if the overlap between their SPP resonances and the black-body spectrum is decreased, we plot the total heat flux between AZO and ITO films across 20 nm gap in Figure 5b as a function of the hot-side temperature, and compare them to the values achievable with SiC films of optimized thickness (20 nm). The cold side temperature was fixed at 300 K. The results presented in Figure 5b demonstrate that the plasmonic thin films outperform SiC films across the whole temperature range, but their advantage becomes more pronounced at higher temperatures, where the SPP-blackbody spectra overlap is maximized.

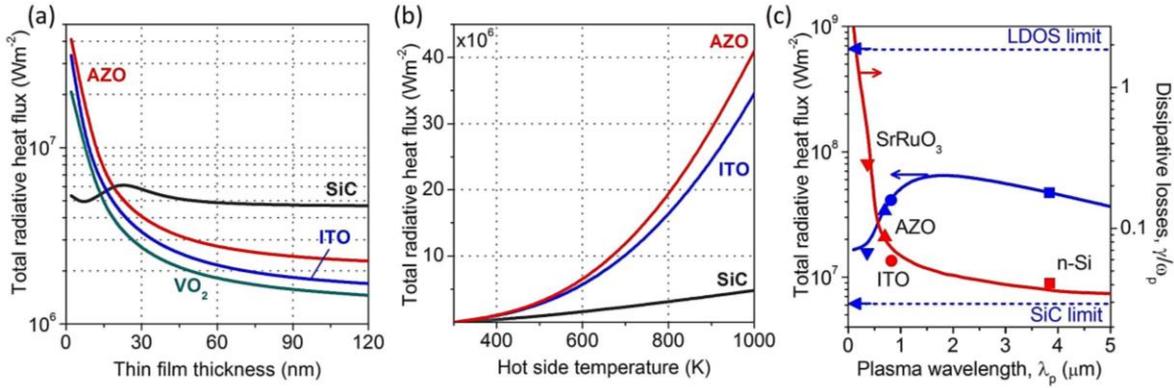

**Figure 5.** (**a**) The total spectrally integrated heat flux between thin films of SiC and plasmonic materials (AZO, VO$_2$) for varying thicknesses. The hot and cold side temperatures are 1000 K and 360 K, respectively. (**b**) Total integrated heat flux between 2 nm-thick plasmonic films (AZO, ITO) and 20 nm-thick SiC films as a function of the hot side temperature. The cold side temperature is 300 K. (**c**) Red line (right vertical axis): optimized level of dissipative losses ($\gamma/\omega_p$) for a Drude metal that maximizes the near-field heat flux between 2 nm-thick plasmonic films as a function of the plasma frequency of the metal with $e_\infty = 3$; Blue line (left vertical axis): the corresponding maximum total heat flux between metals with optimized dissipative losses. The dissipative loss parameters and the total heat flux between 2 nm-films of real plasmonic materials are shown as symbols (downward-pointing triangles: SrRuO$_3$ [42], upward-pointing triangles: AZO, circles: ITO, squares: doped Si [31–33,46,47]). The maximum possible total heat flux across the 20 nm gap (LDOS limit) and the total flux between the SiC films of optimized thickness (SiC) limit are also plotted as dotted lines for comparison. In all the panels, the emitter and absorber are separated by a 20 nm gap.

We now explore in more detail how the extent in the displacement of the SPP resonances from the peak of the blackbody spectrum degrades the efficiency of near-field heat exchange between plasmonic films. We also provide a strategy to compensate for such heat flux reduction with an optimization of the material dissipative losses. In Figure 5c, we plot the dissipative losses ($\gamma/\omega_p$) that maximize the near-field heat exchange between 2 nm-thick films of Drude metals as a function of the metal plasma frequency. The corresponding maximum value of the total heat flux is also shown in this figure. As may be expected, the total heat flux function peaks for the plasma frequencies that provide the best overlap of the SPP-induced peak (calculated via Equation (5)) with the blackbody spectrum at 1000 K. In this case, low dissipative losses are optimal to provide the highest heat flux, as previous asymptotic predictions have shown [49].

Figure 5c also shows that the maximum total heat flux is reduced gradually for materials with plasma frequencies red-shifted from the blackbody emission peak (2.9 μm), and more rapidly for those with blue-shifted plasma frequencies. Note that for materials with progressively shorter plasma wavelengths, the SPP resonance blueshifts and overlaps less with the thermal emission spectrum at a fixed temperature. Our data reveal that increase of the material dissipative losses helps to partially compensate for the flux reduction due to the spectra displacements, with progressively higher dissipation required for materials with plasma frequencies in the visible and ultra-violet frequency bands, in agreement with previous predictions for the case of the non-resonant near-field coupling between lossy metals [7]. On the other hand, for materials with progressively longer plasma wavelengths, the SPP resonance redshifts, and while still overlapping the thermal emission spectrum, gradually moves away from its peak. Thus, progressively lower dissipation losses are required to boost the peak flux value and compensate for the reduced flux bandwidth. It should be noted that we did not include in the above calculations the changes to the material dissipative properties due to electron boundary scattering in thin films. This and other additional factors that may affect the radiative heat flux between thin films are outlined in the discussion section.

The corresponding data for real IR plasmonic materials are also plotted in Figure 5c as symbols. Interestingly, when comparing the data for ITO, AZO, and doped Si with the Drude model prediction, these materials fall on the optimal Drude metal line. On the other hand, the dissipative losses in a recently introduced new plasmonic material $SrRuO_3$ are lower than the optimal Drude metal at its plasma frequency, which results in the maximum heat flux achievable for this material falling below the optimum line in Figure 5c. We did not plot the data for $VO_2$ metal films because this material is approximated with $\varepsilon_\infty = 9$ rather than $\varepsilon_\infty \approx 3$ for ITO, AZO and $SrRuO_3$, which will render any comparison uninformative. The plots for the maximum heat flux achievable with ultrathin plasmonic films in Figure 5c are compared to the flux between SiC films of the optimized thickness, labeled as SiC level, as well as to the maximum possible total heat flux across the 20nm-wide vacuum gap, labeled as the LDOS limit. The ultimate upper limit for near-field heat transfer between two media only depends on the gap width and the hot and cold side temperatures as follows [71]:

$$Q_{max} = \frac{\kappa_B^2}{6\hbar d^2}\left(T_H^2 - T_C^2\right) \tag{6}$$

which scales as $d^{-2}$ with the vacuum gap width. However, the plots of SPP-mediated heat flux as a function of the vacuum gap in Figure 5c feature regions of different power law scaling, including $d^{-3}$ and $d^{-4}$ (see also supplementary Figure S2). This situation is similar to prior calculations of the heat transfer between thin films of polar dielectrics [72]. It can also be seen from Figure 5c, that while ultrathin plasmonic films with the optimized level of dissipative losses outperform the optimally-thick SiC films in a wide range of plasma frequency values, they still fall short by an order of magnitude of reaching the ultimate LDOS limit. This suggests there is still room for further optimization in terms of both material and morphology of the near-field emitters and absorbers.

## 5. Beyond the Single Film: the Effect of Substrate and Multi-Layer Coupling

The data presented in Figures 3–5 corresponded to the case of thin films suspended in vacuum. In most practical devices, however, the thin films will be deposited on a substrate. A full-system optimization for specific device configurations will be dependent on many design considerations, some not necessarily related to the near-field heat transfer process. As a result, the optimum material combinations of the substrate and thin film will vary from case to case. However, some general substrate-induced physical effects that can affect the film-to-film heat exchange need to be discussed.

Our calculations show that in the ideal case, substrates consisting of a low refractive index dielectric with little to no dissipative losses are ideal for maximizing heat flux. However, most dielectrics are not transparent across the wide wavelength range over which energy transfer occurs. Furthermore, many of them support SPhP modes that overlap with the broad SPP resonances of the plasmonic films. Within the frequency band where the substrate SPhP modes and the film SPP modes co-exist, the condition for the SPP propagation on the film-substrate interface is no longer fulfilled. As a result, in this wavelength range the heat flux is only carried by the SPP mode propagating along the film–vacuum interface. This leads to the appearance of local minima in the heat flux spectrum, as observed in Figure 6a for the case of a 2 nm-thin ITO film supported by a $SiO_2$ substrate (red line). The two minima observed in this spectrum correspond to the excitation of two SPhP modes in silica at wavelengths of about 8 μm and 20 μm, which results in the quenching of the SPP-mediated heat flux. Figure 6b shows the corresponding heat transfer function distribution in the energy-momentum space.

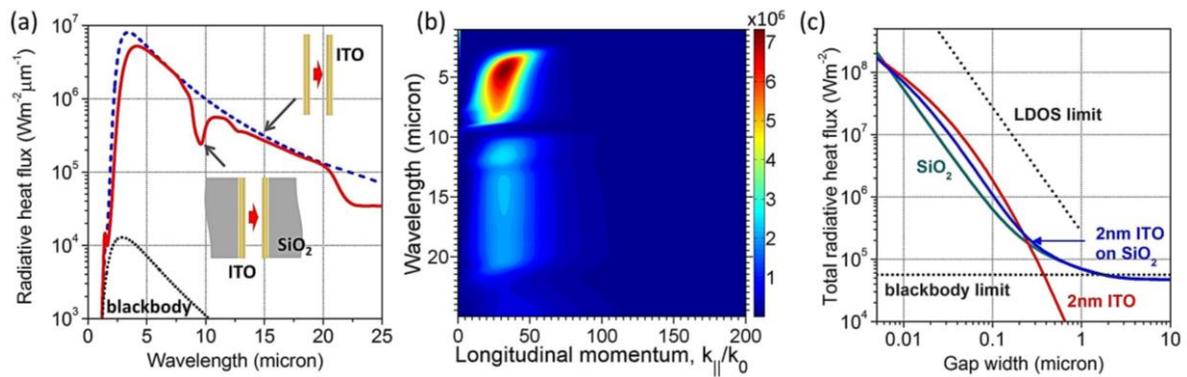

**Figure 6.** (**a**) The heat flux spectra as a function of wavelength between 2 nm-thin suspended ITO films (blue dashed line) and 2nm-thin ITO films supported by $SiO_2$ substrates (red solid line). The emitter and absorber are separated by a 20 nm gap. (**b**) Transmission coefficients from Equation (2) as a function of wavelength and the normalized parallel wave vector for ITO films on $SiO_2$ substrate. (**c**) Total integrated heat flux between 2 nm-thin suspended ITO films (red solid line), 2 nm-thin ITO films supported by $SiO_2$ substrates (blue solid line), and bulk $SiO_2$ substrates (teal solid line). The maximum possible total heat flux across the gap (LDOS limit) and the blackbody limit are shown for comparison (gray dotted lines). The hot and cold side temperatures are 1000 K and 300 K.

Total heat flux integrated over the frequency spectrum as a function of the vacuum gap width is shown in Figure 6c. It reveals subtle differences between the SPhP-mediated and SPP-mediated near-field radiative heat flux enhancement. In particular, the plot for the SPhP-mediated flux between SiO$_2$ surfaces demonstrates the same $d^{-2}$ trend as the maximum possible total heat flux corresponding to the LDOS limit defined by Equation (6). The plots for the SPP-mediated heat flux between thin films deviate from this trend, as in this case the flux is enhanced over a broad range of photon frequencies and momenta, and is not dominated by a contribution from a single narrow peak that reaches very high $k$ values (compare Figure 4a,e with Figure 4b,f). In the limit of the vanishing vacuum gap width, the SPhP-mediated heat flux surpasses the SPP-mediated one, owing to the very strong contribution from the high-momentum photons at such short distances. However, in the more practically-feasible range of 20–300 nm gap widths, plasmonic thin films outperform polar dielectrics (Figure 6c). Since the LDOS limit is only valid in the near-field coupling regime, the corresponding line in Figure 6c was terminated at 1 μm gap width.

We would like to emphasize that the ultrathin plasmonic films considered in this work are almost completely transparent to external far-field illumination [73]. This explains the rapid drop of the plot corresponding to the heat flux between suspended 2 nm-thin ITO films in Figure 6c to the level below the blackbody limit at large gap widths. As the coupling transitions from the near-field to the far-field regime, a larger portion of the emitted light passes through the absorber. However, our calculations reveal that in the near-field regime, nearly all the energy emitted by the thin film on the hot side is absorbed by the thin film on the cold side instead of being transmitted through the film into the vacuum or transparent substrate. Accordingly, various emitter design modifications that rely on reflection and constructive interference of the photons transmitted through the film [45,73,74] for far-field emission or absorption applications are not effective in the near-field coupling regime. Amongst these design modifications, perhaps the most common is a thin emitter/absorber placed onto a Fabry–Perot cavity with a backreflector, which is also known as the Salisbury screen [73,74]. Although this configuration is very useful for enhancing far-field emittance or absorptance, our calculations do not predict any positive effects in the near-field coupling regime.

Previous work has also demonstrated that hyperbolic metamaterials, which are formed by stacking identical metallic thin films sandwiched by dielectric spacers, provide neither enhancement in heat transfer nor spectral tailoring of the heat flux compared to the thin-film case [57]. However, our calculations predict that hybrid thin film stacks composed of polar dielectrics and metals (Figure 7a) can be designed to achieve dramatic spectral shaping with multiple peaks in the spectral radiative heat flux (Figure 7b, red solid line). To achieve SPhP-mediated resonant enhancement rather than quenching of the heat flux, the polar materials should be placed on top of the metal films or substrates. Because polar dielectrics are typically transparent at frequencies where SPhP modes are not supported, multiple dielectrics with SPhP resonances that do not overlap spectrally can be stacked on top of each other to provide multiple peaks in the heat transfer spectrum. If the films are sufficiently thin, the evanescent tails of SPhP modes in intermediate layers can penetrate through the overlying layers and couple to modes on the opposite side of the vacuum gap, thus contributing to near field heat transfer. Since plasmonic metals provide the most broadband peaks in the heat transfer spectrum, they need to be positioned at the bottom of the multilayer stack to avoid screening of narrow-band SPhP modes. The multiple peaks observed in the multilayer stack heat transfer spectrum correspond to the excitation of (starting from the high-energy side) SPP modes in ITO followed by the SPhP modes in $SiO_2$, SiC, $MgF_2$, and $SiO_2$ once again. Figure 7c shows the corresponding heat transfer function distribution for the hybrid multilayer stack in the energy-momentum space. The total integrated heat flux for this structure ($1.75 \cdot 10^7 Wm^{-2}$) exceeds that between $SiO_2$ substrates ($1.33 \cdot 10^7 Wm^{-2}$), but falls below the flux achievable between ITO films, either suspended ($3.43 \cdot 10^7 Wm^{-2}$) or on $SiO_2$ substrates ($2.4 \cdot 10^7 Wm^{-2}$). However, owing to the presence of multiple resonant peaks in the heat flux spectrum, optimized multi-layer structures are expected to provide enhancement over single bulk polar dielectrics, especially in the case of room temperature operation.

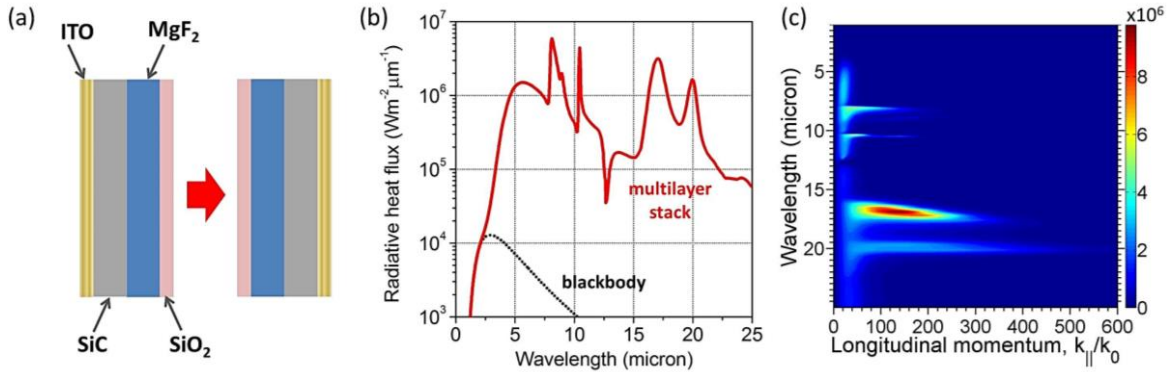

**Figure 7.** (**a**) Schematic of a multi-layer stack composed of 5 nm $SiO_2$ film, 5 nm $MgF_2$ film [63], 5 nm SiC film and 2 nm ITO film. (**b**) The heat flux spectrum as a function of wavelength between two multilayered stacks shown in (a). (**c**) Transmission coefficient from Equation (2) as a function of wavelength and the normalized parallel wave vector for the multilayer stacks. The emitter and absorber are separated by a 20 nm gap, and the hot and cold side temperatures are 1000 K and 300 K, respectively.

Based on the data presented in Figures 6 and 7, thin plasmonic films can either efficiently screen SPhP modes when placed on top of a polar dielectric substrate, or enable SPhP modes to transfer energy when placed below a polar dielectric film. This suggests an interesting application for phase change materials, which can switch from metal to insulator phases under a variety of external stimuli such as temperature or an applied voltage [15,16,44,62,75–77]. One such material is vanadium dioxide, which undergoes an insulator-to-metal phase transition at 68 °C [15,62,78]. The dielectric permittivities extracted from experimental measurements of $VO_2$ films are plotted in Figure 8a for both metal (at 360 K) and insulator (at 341 K) phases.

Figure 8b compares the near-field heat flux spectra between 2 nm-thin $VO_2$ films on $SiO_2$ substrates, with a hot side temperature of 1000 K and a cold side temperature either at 360 K (red dotted line) or 341 K (solid blue line). Clearly, a dramatic change in the heat flux spectra can be observed, which can be attributed to the phase transition of $VO_2$ on the cold side. When the $VO_2$ film on the cold side is in its insulating phase, a SPhP mode will be supported at the interface between the $SiO_2$ substrate and $VO_2$ film, which will couple to the SPP mode of the metallic $VO_2$ film on the hot side. However, when the $VO_2$ film on the cold side is in its metallic phase, only the SPP modes supported by the $VO_2$ films on both the hot and cold sides will couple. As a result, as the $VO_2$ film transitions from its insulating phase to its metallic phase, the spectral heat flux switches from a narrowband peak to a broad spectrum. We would like to emphasize that the switching behavior between very intense peaks observed in Figure 8b stems from the unique near-field coupling and spectral tailoring properties of the ultra-thin films.

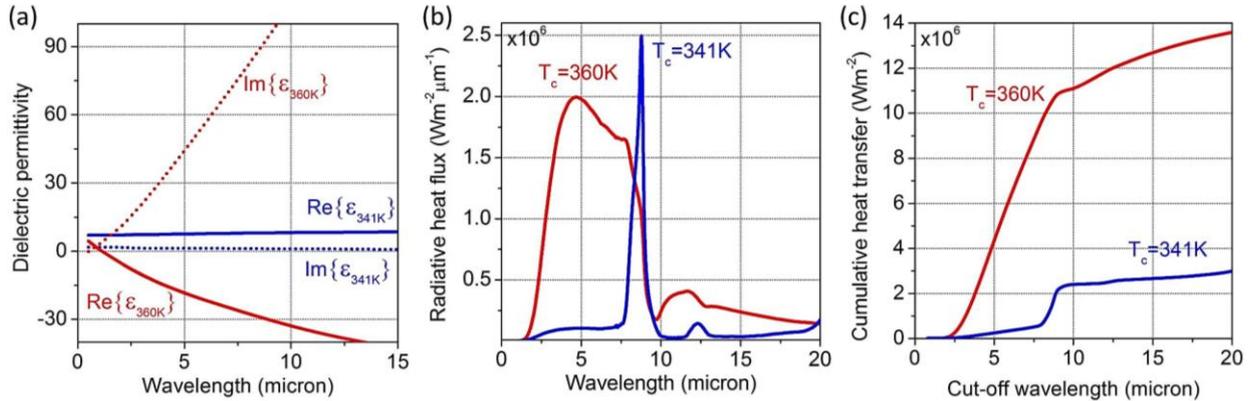

**Figure 8.** (**a**) Complex dielectric permittivities of $VO_2$ in the metal phase (red lines) and insulator phase (blue lines). The real part of permittivity is plotted as solid lines and the imaginary parts as dotted lines. (**b**) The heat flux spectra as a function of wavelength between 2 nm-thin $VO_2$ films on $SiO_2$ substrates at an emitter temperature of 1000 K and an absorber temperature either at 360 K (red dotted line) or 341 K (solid blue line). The emitter and absorber are separated by a 20 nm vacuum gap.

Figure 8c displays the accumulated heat flux transferred by photons with energies above sliding cut-off energy, which shows how much the total flux will change between the two cases compared in Figure 8b. The SPP-mediated heat flux between $VO_2$ films in their metal phase is clearly larger than that between a $VO_2$ film in the metal phase on the emitter side and a $VO_2$ film in the insulator phase on the absorber side. For comparison, previous work on the manipulation of near-field radiative heat transfer between bulk $VO_2$ emitter and absorber at room temperature predicted that the heat transfer between plates in their metal phase was not significant and actually lower compared to the $VO_2$ insulator phase [15]. In the latter case, the enhancement to heat transfer was attributed to the resonant coupling of SPhP polariton modes supported by the insulator phase of $VO_2$ at low frequencies (above 20 μm in wavelength). However, in this work the radiative heat flux is dominated by the strong contribution from the SPP modes on the hot side metal $VO_2$ film and the SPhP modes on cold side $SiO_2$ substrate.

## 6. Summary, Discussion, and Future Outlook

To summarize, we demonstrate through rigorous numerical modeling that ultra-thin plasmonic films dramatically enhance near-field radiative heat transfer across nanoscale vacuum gaps. The film-to-film heat flux is over an order of magnitude higher compared to bulk emitters and absorbers made of the same plasmonic materials. A wide range of new plasmonic materials have been recently synthesized and explored exhibiting plasma frequencies in either red or infrared part of the spectrum, which make these materials promising candidates for near-field heat transfer applications. Many of the unconventional IR plasmonic materials are also attractive candidates for high-temperature applications owing to their high melting temperatures (e.g., 2248 K for AZO, 3203 K for TiN, 3293 K for Ta, 3695 K for W, 1800–2200 K for ITO, and 2240 K for $VO_2$).

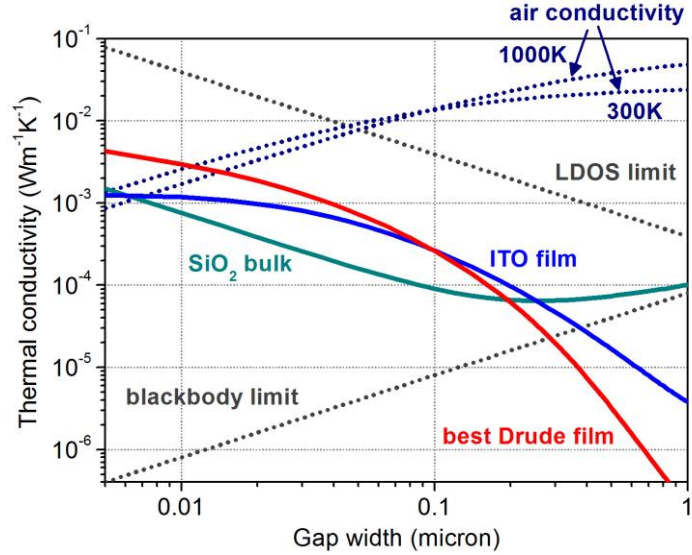

**Figure 9.** Comparison of the thermal conductivity across a narrow gap via thermal radiation through vacuum and via heat conduction through air as a function of the gap width. The solid lines correspond to the conductivity via near field radiation between 2 nm-thin ITO films (blue), bulk $SiO_2$ (teal), and 2 nm-thin plasmonic films with optimized Drude model parameters (red). The dotted gray lines denote the LDOS and the blackbody radiation limits. The dotted dark blue lines correspond to the heat conductivity through air slab for air temperatures of 300 K and 1000 K calculated via Equation (7). The bulk thermal conductivity and viscosity values are taken from the literature: $K_B^{300K} = 0.026\ Wm^{-1}K^{-1}$, $K_B^{1000K} = 0.0675\ Wm^{-1}K^{-1}$, $m^{300K} = 1.8 \times 10^{-5}\ Pa \times s$, $m^{1000K} = 4.1 \times 10^{-5}\ Pa \times s$.

Our calculations predict that total spectrally integrated heat flux between ultrathin plasmonic films exceeds what is achievable with polar dielectrics. However, even the best optimized plasmonic material with the dielectric permittivity described by a Drude model falls short of the ultimate energy transfer limit for a given gap separation as well as hot and cold side temperatures. This is illustrated in Figure 9 (see also Figure 6c), which compares thermal conductivity $K = Q \cdot d /(T_H - T_C)$ via near-field radiative channels for several representative structures, including: 2 nm ITO films, bulk $SiO_2$, and 2 nm plasmonic films with optimized Drude parameters ($l_p = 1.8 mm$, $g/w_p = 0.048$, $\varepsilon_\infty = 3$). The LDOS and the blackbody limits are shown for comparison as gray dotted lines. To put these values into perspective, we also plot the conventional thermal conductivity through the air slab confined in the narrow gap between two surfaces. The effective thermal conductivity of the air slab is reduced from the bulk air conductivity value $K_B$, and its scaling with the slab thickness $d$ can be described via a suppression function $S(\eta)$, $\eta = \Lambda/d$ [79,80]:

$$K = K_B \times S(\eta); \quad S(\eta) = 1 - \frac{3}{4}\eta \times \left(1 - 4E_s(\eta^{-1})\right) \approx \left(1 + \frac{4}{3}\eta\right)^{-1}. \tag{7}$$

where $E_s(x)$ is the exponential integral function [80] and $\Lambda$ is the mean free path of air molecules, which depends on gas temperature $T$, pressure $P$, and viscosity $\mu$. The mean free path of air molecules at room temperature ($\Lambda_0$) equals to 68 nm, and at any given temperature can be calculated as: $\mathsf{L} = \mathsf{L}_0 \times (m/m_0) \times (P/P_0) \times (T/T_0)^{1/2}$ [81]. In Figure 9, we plot the air slab thermal conductivity under constant pressure of 1 atmosphere and for temperatures of 1000 K and 300 K (dotted dark blue lines). It should be noted that for the gaps smaller than the molecule mean free path, thermal conductance of air is ballistic and does not depend on either the mean free path or the separation distance [82]. In this regime, thermal conductivity scales linearly with *d*, as can be clearly seen in Figure 9 for small *d* values.

Although in the far field regime thermal energy flux through conduction always significantly exceeds that through radiation, the data in Figure 9 show that in the case of the SPP-enhanced near-field heat transfer the situation may be reversed. It was previously predicted that the heat flux at the LDOS limit can surpass the heat conduction through air [83], but our data now reveal that thin plasmonic films of optimized Drude materials are capable of such strong radiative enhancement (red solid line). However, comparison of the LDOS limit to the data for the thin film of the Drude metal with optimized parameters suggests that significant room for improvement still exists. The results from the optimization of the film thickness and material presented in this work could help guide future research. In particular, plasmonic materials with dielectric permittivities not fully described by the Drude model can be explored for further optimization of SPP-mediated heat transfer between thin films.

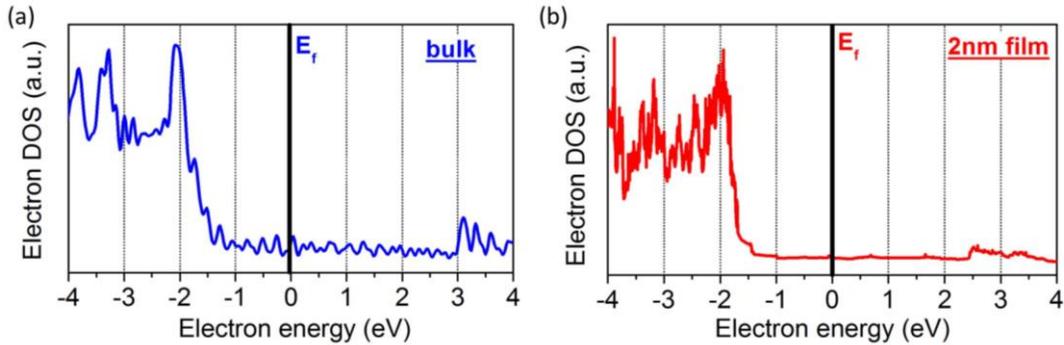

**Figure 10.** The electron density of states for (**a**) bulk gold and (**b**) a 2 nm-thick film of gold.

Since the heat flux enhancement is a result of the SPP mode splitting, the highest flux is predicted in the case of the ultra-thin films with a vanishing thickness. This is different from the case of thin films composed of polar dielectrics, which make use of the same mechanism of surface mode splitting, yet maximize the heat flux at an optimum non-vanishing film thickness. In turn, the heat flux spectral shaping and enhancement mechanism of thin films of both plasmonic and polar dielectrics is completely different from that recently demonstrated for spectral shaping with semiconductor thin films [56]. The latter relies on the photon DOS manipulation owing to excitation of guided modes trapped by the total internal reflection rather than surface polaritons, and also maximizes for non-vanishing film thickness. Deposition of ultrathin continuous plasmonic films is challenging, but successful experimental demonstrations of metal and metal oxide films of few-nm thicknesses can be found in literature [73,84–86].

Due to the low-dimensional morphology of the emitter and absorber, electron quantum confinement effects can modify their SPP characteristics, which in some cases can result in plasmon damping [87,88]. Our preliminary calculations of the electron bandstructure indeed reveal some differences in the density of electron states (e-DOS) in bulk and 2 nm-thin film of gold (Figure 10). The e-DOS calculation was performed by using the QUANTUM ESPRESSO simulation package [89] with the generalized gradient approximations for the exchange-correlation energy [90] and the projector augmented wave method [91]. Band structures are obtained first using Gaussian broadening method, which are then interpolated onto a much finer mesh with the tetrahedra method to calculate the e-DOS [92]. The comparison of the e-DOS energy spectra for bulk (Figure 10a) and thin-film (Figure 10b) shows some significant changes in the higher-energy part of the spectrum. However, e-DOS values in the low-energy range relevant for coupling with the infrared SPP modes contributing to the resonant heat transfer do not vary significantly between bulk and film. Nevertheless, for more accurate design of ultrathin film near-field emitters and absorbers, quantum-confinement-corrected models (and/or experimental measurements) of their dielectric permittivities will be required.

We also demonstrated that the radiative heat flux spectrum can be further shaped by the choice of the film-substrate combination, especially in the case of substrates supporting SPhP modes. The film-substrate interactions can also include hybridization of the plasmon modes with the bulk optical phonon modes, excitation of acoustic phonons in the films, and strain resulting from the lattice mismatch between thin film and substrate [87]. These effects may play different roles for films of different materials and need to be considered on the case-by-case basis. On the other hand, the use of phase-change materials as either thin films or substrates allows for the dynamical switching of the heat flux spectrum between SPP-mediated and SPhP-mediated peaks. Other types of phase-change or voltage-tunable materials that can be explored in addition to $VO_2$ studied in this work including ferroelectrics [16], W-$VO_2$ alloys [93], ITO [77], and $Ge_2Sb_2Te_5$ (GST) [94] just to name a few.

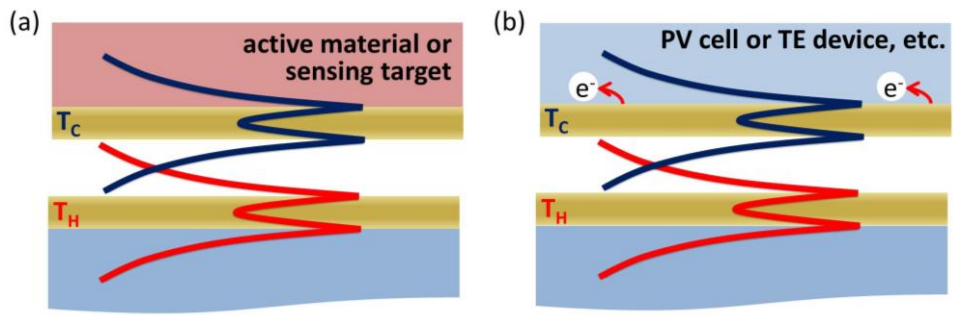

**Figure 11.** Schematics of possible heat-pumped SPP-mediated platforms for (**a**) Sensing, spectroscopy, LED, PV, and (**b**) hot-electron cells, catalysis or thermoelectrics.

The ability to tune the plasma frequency in plasmonic materials [39] makes the SPP-mediated rather than SPhP-mediated near-field heat transfer platforms promising candidates for heat-activated infrared near-field sensing, spectroscopy, fluorescence enhancement, and TPV applications as schematically illustrated in Figure 11a. The additional advantage offered by the plasmonic films is the hybrid nature of their surface plasmon polariton modes, which are in fact resonant collective oscillations of free charge carriers. This provides an additional pathway for harvesting heat energy of the plasmonic

emitter—via hot electron tunneling from the thin metal absorber to the substrate material as shown in Figure 11b. Hot electrons cool fast due to scattering by phonons, on picosecond timescale in most metals, and need to be extracted before they cool down. The ultrathin films considered in this work, however, have thicknesses far below the electron mean free path in metals, and thus offer a high probability for the hot electrons to be extracted before they lose their energy to lattice vibrations. Furthermore, high electron-plasmon scattering rates provide a mechanism for 'hot electron protection,' shielding free hot carriers from losing their energy through electron–phonon scattering [95]. Creation of high-quality single-crystalline films will nevertheless be essential to reduce the electron–photon scattering rates in metals and to improve the efficiency of the hot carriers extraction [96]. Overall, this SPP-mediated energy-harvesting mechanism can form the basis for new types of photovoltaic [97], catalytic [98–100], thermoelectric [101], or imaging [102] platforms pumped by heat rather than light.

**Acknowledgments**

This work has been supported by the US Department of Energy, Office of Science, and Office of Basic Energy via 'Solid State Solar-Thermal Energy Conversion Center (S3TEC)', Award No. DE-SC0001299/DE-FG02-09ER46577 (for solar thermal applications) and Grant no. DE-FG02-02ER45977 (for near-field transport and photon density of states tailoring via confinement effects).

**Author Contributions**

S.V.B. conceived the original idea and led the manuscript writing. J.K.T., Y.H., and S.V.B. developed the radiative heat transfer model. V.C., J.K.T. and S.V.B. performed comparison to thermal conductivity of air. J.Z. performed calculations of the electron density of states. G.C. and S.V.B. supervised the research. All authors contributed to scientific discussions and manuscript writing and editing.

**Conflict of Interest**

The authors declare no conflict of interest.